\documentstyle[emuapj_acc,apjfonts,epsf]{article}

\def\lapp{\ifmmode\stackrel{<}{_{\sim}}\else$\stackrel{<}{_{\sim}}$\fi}
\def\gapp{\ifmmode\stackrel{>}{_{\sim}}\else$\stackrel{<}{_{\sim}}$\fi}

\begin{document}

\title{Timing of Millisecond Pulsars in NGC~6752: \\ Evidence for
a High Mass-to-Light Ratio in the Cluster Core}

\author{N.~D'Amico,\altaffilmark{1,2}
A.~Possenti,\altaffilmark{1}
L.~Fici,\altaffilmark{3}
R.~N.~Manchester,\altaffilmark{4} 
A.~G.~Lyne,\altaffilmark{5}
F.~Camilo,\altaffilmark{6}
and J.~Sarkissian\altaffilmark{7}}
\medskip

\affil{\altaffilmark{1}Osservatorio Astronomico di Bologna,
Via Ranzani 1, 40127 Bologna, Italy}
\affil{\altaffilmark{2}Osservatorio Astronomico di Cagliari,
Loc. Poggio dei Pini, Strada 54, 09012 Capoterra (CA), Italy}
\affil{\altaffilmark{3}Universita di Bologna, Dipartimento di Astronomia,
Via Ranzani 1, 40127 Bologna, Italy}
\affil{\altaffilmark{4}Australia Telescope National Facility,
CSIRO, PO Box 76, Epping, NSW 1710, Australia}
\affil{\altaffilmark{5}University of Manchester, Jodrell Bank
Observatory, Macclesfield, Cheshire SK11~9DL, UK}
\affil{\altaffilmark{6}Columbia Astrophysics Laboratory, 
Columbia University, 550 West 120th Street, New York, NY~10027}
\affil{\altaffilmark{7}Australia Telescope National Facility,
CSIRO, Parkes Observatory, PO Box 276, Parkes, NSW 2870, Australia}

\bigskip

\begin{abstract}
Using pulse timing observations we have obtained precise parameters,
including positions with $\sim 20$ mas accuracy, of five millisecond
pulsars in NGC~6752.  Three of them, located relatively close to the
cluster center, have line-of-sight accelerations larger than the
maximum value predicted by the central mass density derived from
optical observation, providing dynamical evidence for a central
mass-to-light ratio $\gapp 10$, much higher than for any other
globular cluster. It is likely that the other two millisecond pulsars
have been ejected out of the core to their present locations at 1.4
and 3.3 half-mass radii, respectively, suggesting unusual non-thermal
dynamics in the cluster core.
\end{abstract}


\section{Introduction}

During a deep search of the globular cluster (GC) system for
millisecond pulsars (MSPs), carried out using the Parkes radio
telescope, we discovered (D'Amico et al. 2001a)\nocite{dlm+01} a
binary MSP (hereafter PSR-A) associated with NGC~6752, a
core-collapsed cluster showing evidence of mass segregation (Ferraro
et al. 1997)\nocite{fcbro97}. The resulting knowledge of the dispersion
measure (DM) for this cluster has facilitated the discovery of four
additional MSPs (hereafter PSRs B, C, D, E) in the same cluster
(D'Amico et al. 2001b)\nocite{dpm+01}. We have made frequent
observations using the Parkes telescope to obtain coherent timing
solutions for these pulsars.  These results can be used to estimate a
variety of physical properties of the pulsars and of the host cluster.
In particular, if the measured period derivatives are dominated by the
dynamical effects of the cluster gravitational field, they can be used
to constrain the mass-density distribution of the cluster, giving
information on the GC's dynamical status and on the population of
optically unseen cluster members.  In this paper we report on the
first 15 months of pulse timing observations, providing rotational and
positional parameters for the five MSPs known in NGC~6752, and we use
the pulsar positions and inferred accelerations to probe the dynamical
status of the cluster.

\section{Observations and Results}
Regular pulsar timing observations of NGC~6752 have been carried out
since 2000 September with the Parkes 64-m radio telescope, using the
center beam of the multibeam receiver at 1400 MHz. The hardware system
is the same as that used in the discovery observations (D'Amico et al.
2001a)\nocite{dlm+01}.  The effects of interstellar dispersion are
minimized by using a filter bank having 512 $\times$ 0.5 MHz frequency
channels for each polarization. After detection, the signals from
individual channels are added in polarization pairs, integrated, 1
bit-digitized every 125 $\mu$s, and recorded to magnetic tape for
off-line analysis. Observation times are typically 1--2 hours.
Pulse times of arrival (TOAs) are determined by fitting a standard
high signal-to-noise ratio pulse profile to the observed mean pulse profiles
and analyzed using the program {\sc
tempo}\footnote{See http://www.atnf.csiro.au/research/pulsar/timing/tempo.} and
the DE200 solar system ephemeris (Standish 1982)\nocite{sta82}.

Table~1 lists the timing parameters obtained for the five pulsars,
including precise positions.  Because these pulsars have relatively
low DMs$\sim$33 cm$^{-3}$pc, interstellar 
scintillation strongly
affects their detectability.  In particular, PSR-B was detected on
only 27 of 140 observations of the cluster.  The mean flux densities
at 1400 MHz ($S_{1400}$) in Table~1 are averaged values, derived from
the observed signal-to-noise ratios. Non-detections were accounted for
by assuming $S_{1400}$ values corresponding to half
the detection limits. As can be seen from Table~1, all but PSR-A are
isolated MSPs. PSRs B, D and E are located close to the cluster
center.  PSR-D has the third largest period derivative, $\dot{P} = 9.6
\times 10^{-19},$ among known MSPs, after PSR B1820$-$30A in NGC~6624 and
PSR B1821$-$24 in M28, suggesting that the $\dot{P}$ is dominated by
the line-of-sight acceleration in the cluster gravitational
field. This interpretation is supported by the large negative
$\dot{P}$ values observed for PSR B and E, which are also located close to
the cluster center.

PSR-C is located $2\farcm7$ from the cluster center, equivalent to
about 1.4 half-mass radii or 24 core-radii, assuming a core radius
$\theta_{c}=6\farcs7$ (Lugger, Cohn \& Grindlay 1995), a half-mass
radius $\theta_{hm}=115\arcsec$ (Djorgovski 1993)\nocite{d93}, and an
optical center for the cluster of (J2000) R.A.=19$^{\rm h}$ 10$^{\rm
m}$ $51\fs8$ and Dec.=$-$59$^\circ$ 58$\arcmin$ 55$\arcsec$ (Harris
1996). Previously, the largest offset of an associated pulsar from a
GC center was for PSR B2127+11C (Prince et al. 1991)\nocite{pakw91}, a
member of a double-neutron-star eccentric binary in M15, located at
$\sim$ 0.8 half-mass radii.  Given the large offset of PSR-C from the
GC center, its period derivative is not significantly affected by the
GC potential well, allowing the measurement of the characteristic age
($\tau_{\rm c} = P/2\dot{P} = 3.8\times10^{10}$ yr, the largest
among known pulsars), the surface magnetic field ($B_{\rm surf} = 3.2\times
10^{19}(P\dot{P})^{1/2} = 1.1\times10^{8}$ Gauss) and the rotational
energy loss ($\dot{E}$ = 3.95$\times$ 10$^{46}\dot{P}/P^3$
=5.9$\times10^{32}$ erg s$^{-1}$).

PSR-A, the first pulsar discovered in this cluster, is located even
farther from the center, at $6\farcm4$, equivalent to 3.3 half-mass
radii or $\sim 57$ core radii. Given this large radial offset, one
could question the association of the pulsar with the cluster. Based
on the 19 MSPs with $S_{436}\gapp 2$ mJy detected by the Parkes
Southern Pulsar Survey (Lyne et al.  1998)\nocite{l98} and assuming a
typical spectral index $\sim-1.9$ (Toscano et al. 1998)
\nocite{tbms98} for MSPs, the probability of chance superposition of a
Galactic field MSP with $S_{1400}\gapp~0.2$ mJy within $6\farcm4$ of
the center of NGC~6752 is at most $10^{-5}$.
This probability is further reduced by noting that all five pulsars
have very similar DM values.   Also in this case the observed
period derivative can be used to measure the pulsar parameters
$\tau_{\rm c} = 1.7\times10^{10}$ yr,
$B_{\rm surf} = 1.0\times10^{8}$ Gauss,
and $\dot{E} =5.9\times10^{32}$ erg s$^{-1}$.

Accurate DMs were obtained for each pulsar by splitting the 256-MHz
bandwidth into two adjacent sub-bands and computing the differential
delays.  The maximum deviation ($\sim$0.3 cm$^{-3}$ pc) of the DM
values derived for each pulsar from the average (33.36 cm$^{-3}$ pc)
is similar to that observed in other GCs hosting several MSPs like
47~Tucanae (Freire et al. 2001) \nocite{fkl+01} and M15 (Anderson et
al.  1990)\nocite{agk+90}. Given the wide angular offsets of some of
the pulsars, the scatter in some of the DMs could arise from gradients
in the Galactic column density across different lines-of-sight toward
the cluster (Armstrong, Rickett \& Spangler 1995)\nocite{ars95}. It
could also arise from an enhanced electron density within the cluster
(see Freire et al. 2001). Excluding the peripheral PSR-A, a rough
estimate of the density of such gas is given by
$n_e\sim\langle(\Delta{\rm
DM})^2\rangle^{1/2}/(D\langle\theta_{\perp}^2\rangle^{1/2}) =
0.025\pm0.005$ cm$^{-3}$, where
$\langle(\Delta\mbox{DM})^2\rangle^{1/2}$ is the rms deviation in the
DMs and $\langle\theta_{\perp}^2\rangle^{1/2}$ is the one-dimensional
dispersion of the angular offsets in radians with respect to the GC
center in the plane of the sky for the 4 inner MSPs. $D=4.1\pm 0.1$
kpc (Renzini et al. 1996) is the cluster distance (the errors are
reported at the 1$\sigma$-level, as everywhere in this paper and in
Table~1).  This $n_e$ value is about a factor of ten smaller than that
inferred from the four MSPs (with $P<10$ ms) in the core-collapsed
cluster M15 and $\sim 40\%$ of that derived using a more refined
modeling of the plasma content in 47~Tuc (Freire et al. 2001).

NGC 6752 was observed recently for $\sim 28700$ s with the ACIS-S
detector aboard the {\it Chandra} X-ray observatory by Pooley et
al. (2002). The position of PSR-D is consistent with that of the {\it
Chandra} source labeled CX11 by Pooley et al. All the MSPs but PSR-A
are located in the {\it Chandra} field of view, but PSR-C is outside
the half-mass radius region searched for X-ray sources by Pooley et
al. Thus we have processed the full ACIS-S3 image in the 0.5--6.0 keV
band using the CIAO 2.2 software{\footnote{See
http://asc.harvard.edu/ciao .}}, resulting in the detection of two
additional probable X-ray counterparts to the MSPs.  Using the {\it
wavedetect} tool, we have found a weak soft source whose error circle
encloses PSR-C, having a hardness ratio (as defined by Grindlay et
al. 2001a) $>$ 5.5. Assuming isotropic X-ray emission and a hydrogen
column density $N_H=2.2\times 10^{20}$ cm$^{-2}$, we
obtain{\footnote{See http://heasarc.gsfc.nasa.gov/Tools/w3pimms.html .}}  
for three different spectral models (a power-law of photon index
$-2.5$, a blackbody with $kT=0.3$ keV or a thermal bremsstrahlung with
$kT=1$ keV) similar values of the X-ray luminosity in the 0.5--2.5 keV
band, $L_{X}$=2.2$\times10^{30}$ erg s$^{-1}$, corresponding to a
conversion efficiency $L_{X}/\dot{E}\sim 0.004$.  This is somewhat
higher than that predicted on the basis of both the sample of MSPs
observed in 47 Tuc and the sample of MSPs in the Galactic field
(Grindlay et al. 2002\nocite{ghe+02}).  Using the {\it celldetect}
tool, we have also found marginal evidence for a slightly harder
source compatible with the position of PSR-B, having a hardness ratio
$\sim$ 4 and $L_{X}$ = 1.1$\times10^{30}$ erg s$^{-1}$. This source is
surrounded by many other sources and the nominal {\it celldetect}
signal-to-noise ratio, $\sim$1.5, is low.  No X-ray source is
associated with PSR-E, with an upper limit to the X-ray luminosity in
the 0.5--2.5 keV band of $10^{30}$ erg s$^{-1}$.  Based on the three
probable positional coincidences, we have used the MSP positions
obtained via radio timing to derive a corrected {\it Chandra}
astrometric solution, requiring a shift of the {\it Chandra} positions
by $-0\farcs05$ in RA and $-0\farcs03$ in Dec, with a final rms (of
the differences between {\it Chandra} and radio positions) of
$0\farcs28$ in RA and $0\farcs39$ in Dec, consistent with the expected
positional uncertainties of $\sim 0\farcs3$ (Pooley et al. 2002).

Figure~1 shows an optical image of the cluster, the
positions of the five MSPs and the contour levels of the {\it Chandra}
image.  Pooley et al. (2002) pointed out that the X-ray sources in the
central $1\farcm6\times 1\farcm6$ region lie in the south-east
quadrant. While they ascribe this to chance, we note that most of the
X-ray sources within the half-mass radius region are roughly
distributed along a stretched S-shaped pattern, whose elongated ends
are oriented in the east--west direction.

\section{Central Mass-to-Light Ratio}

The large negative $\dot{P}$ values observed in PSRs B and E can be
used to derive lower limits to the line-of-sight accelerations,
$a_{l}/c=\dot{P}/P = -9.6\pm 0.1\times 10^{-17}$ s$^{-1}$, for both
pulsars, which are the largest known after those of PSRs~B2127+11A and
D in the core of M15 (Anderson et al.  1990)\nocite{agk+90}.
Contributions from centrifugal acceleration (Shklovskii 1970),
differential Galactic rotation (Damour \& Taylor 1991) and vertical
acceleration in the Galactic potential (Kuijken \& Gilmore 1989) are
all negligible. Finally, according to Figures~3 and 4 of Phinney
(1993), the probability that the accelerations of both PSRs B and E
are dominated by a nearby star in the cluster is $\lapp 10^{-4}$.
Thus we can conclude that the inferred high values of
$|a_{l}/c|$ are due to the potential well of NGC~6752.

A lower limit to the mass-to-light ratio in the inner regions
of NGC~6752 can be derived from the following rule, which holds to
within $\sim$10\% in all plausible cluster models (Phinney 1992\nocite{phi92}):
\begin{eqnarray*}
 & \left|\frac{\dot{P}}{P}(\theta_{\perp})\right| < 
\left|\frac{a_{l,max}(\theta_{\perp})}{c}\right| \simeq & \\
\simeq & 1.1\frac{G}{c}\frac{M_{cyl}(<\theta_{\perp})}{\pi
D^2\theta^2_{\perp}}=5.1\times 10^{-18} \frac{\cal{M}}{{\cal
L}_V}\left(\frac{\Sigma_{V}(<\theta_{\perp})}{10^4~{\rm L_{V\odot}
pc^{-2}}}\right){\rm s^{-1}} & .
\label{eq:Sigma}
\end{eqnarray*}
Here $\Sigma_{V}(<\theta_{\perp})$ is the mean surface brightness
within a line of sight subtended by an angle $\theta_{\perp}$ with
respect to the cluster center, $M_{cyl}(<\theta_{\perp})$ is the mass
enclosed in the cylindrical volume of radius
$R_{\perp}=D\theta_{\perp}$ and $\cal{M}/{\cal L}_V$ is the mean {\it
projected} mass-to-light ratio in the V-band. 
{
\vskip 0.5truecm 
\epsfxsize=8.9truecm 
\epsfysize=6.2truecm
\epsfbox{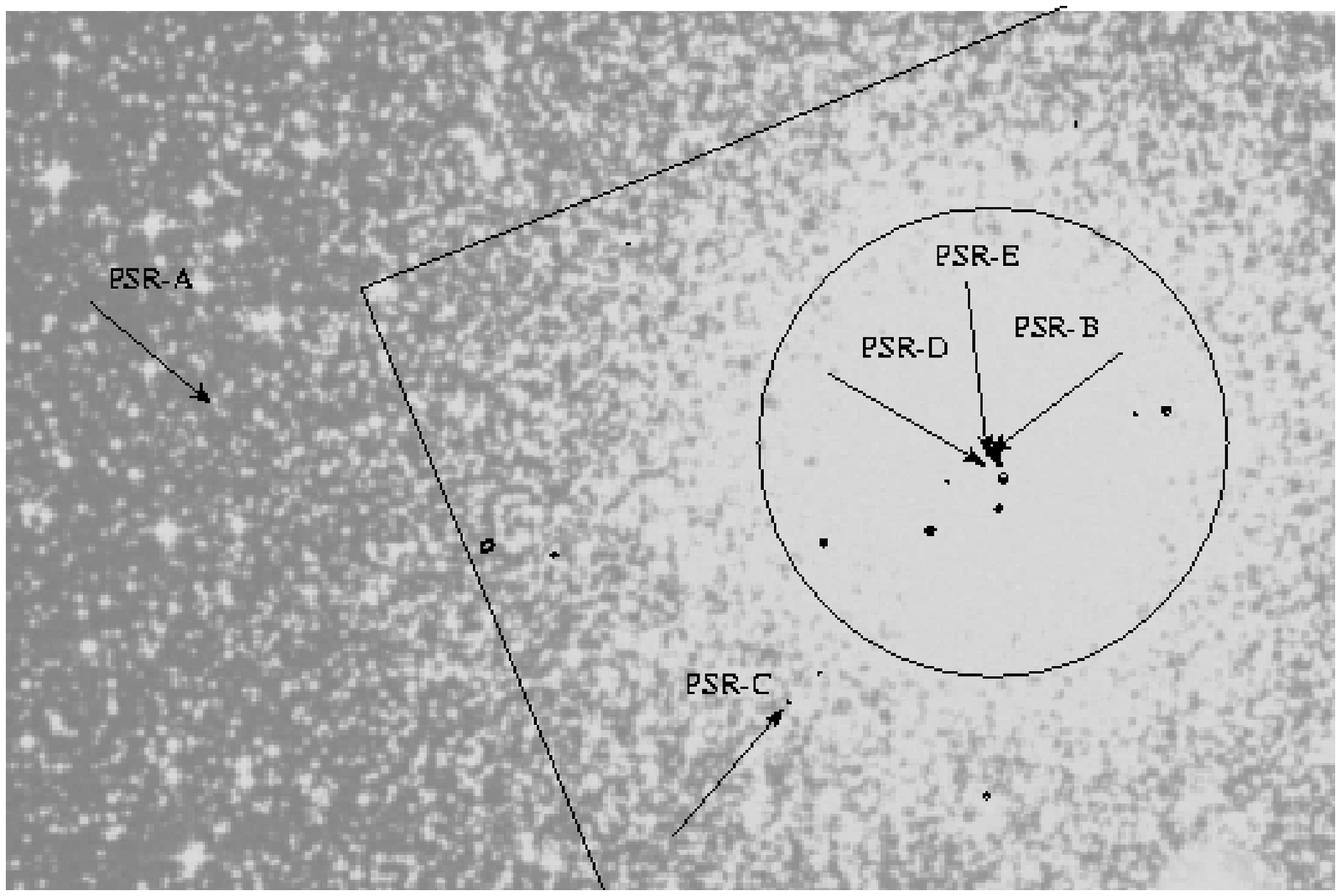} 
\figcaption[fig1.eps]{\label{fig:fig1}
\small{Positions of the 5 millisecond pulsars
(arrows) and of the X-ray sources (black contours) in NGC~6752,
superimposed on an optical image retrieved from the Digital
Sky Survey. The circle indicates the half-mass radius region
($\theta_{hm}=115\arcsec$). 
The rectangular box indicates the {\it Chandra}
ACIS-S3 field of view. North is to the top and east is to the left.}}
\vskip 0.5truecm
}

We use the most recent
published brightness profile for this cluster (Lugger, Cohn \&
Grindlay 1995\nocite{lcg95}), normalized to the central surface
brightness in the V-band reported by Djorgovski (1993), in order to
plot the curves of maximum $|a_l/c|$ for different values of
$\cal{M}/{\cal L}_V$. 
In Figure~2 the histogram represents the data
for $\cal{M}/{\cal L}_V$=1.1, as suggested by Pryor \& Meylan
(1993). This greatly underestimates the observed values of
$\dot{P}/P$.  The dashed lines are analytical fits (with reduced
$\chi^2 \sim$1) to the observed data, scaled according to increasing
values of $\cal{M}/{\cal L}_V$. Only $\cal{M}/{\cal L}_V\gapp$9 can
account for the observed $|\dot{P}/P|$ of PSRs B and E. An even larger
$\cal{M}/{\cal L}_V\gapp$ 13 is required if PSR-D has a negligible
intrinsic $\dot{P}/P,$ as might be expected given the close projected
positions and the similar absolute values of $\dot{P}/P$ for PSRs B, D
and E. A reasonable interpretation is that PSR D is at about the same
distance from the GC center as the PSRs B and E, but in the closer
half of the cluster, whereas PSRs B and E reside in the further half.
In this case, for these 3 MSPs the intrinsic $\dot{P}/P\sim 5\times
10^{-18}$ can be estimated as the average of those of PSRs B (or E)
and D.  Values of $\cal{M}/{\cal L}_V$ in the interval $9-13$ result
if the observed scalings between X-ray luminosity and spin-down power
for MSPs are used to estimate the magnetic-dipole braking contribution
to $\dot{P}/P$ for PSR-D (see caption of Fig.~2).  We finally note
that these estimates of the projected $\cal{M}/{\cal L}_V$ are
independent of distance, excepting the effects of extinction, which
are very small for NGC~6752, with E(B$-$V) = 0.04 (Harris
1996)\nocite{h96}, and of modeling of the cluster potential.

These results imply (see eq. \ref{eq:Sigma}) that there must be
$M_{cyl}(<\theta_{\perp,E})\gapp 1.3\times 10^4~{\rm M_\odot}$ of
matter in the form of low-luminosity stellar objects within the
projected radius of PSR-E, $0.15$ pc. Adopting the prescription of
Djorgovski (1993) and a core radius $\theta_c=6\farcs7$
(Lugger, Cohn \& Grindlay 1995) this in turn corresponds to a central
mass density $\gapp 7\times 10^5~{\rm M_\odot}$pc$^{-3}$, at least
five times larger than that derived from measurements in the optical
band (Pryor \& Meylan 1993). The stars whose initial mass was in the
interval 0.6--0.8~${\rm M_\odot}$ (bracketing the current turn-off
point of the cluster) now dominate the total integrated V-band
luminosity of NGC~6752, $1.2\times 10^5~{\rm L_{V\odot}}$ (Djorgovski
1993)\nocite{d93}. 

{
\vskip 0.5truecm \epsfxsize=8.5truecm \epsfysize=9.truecm
\epsfbox{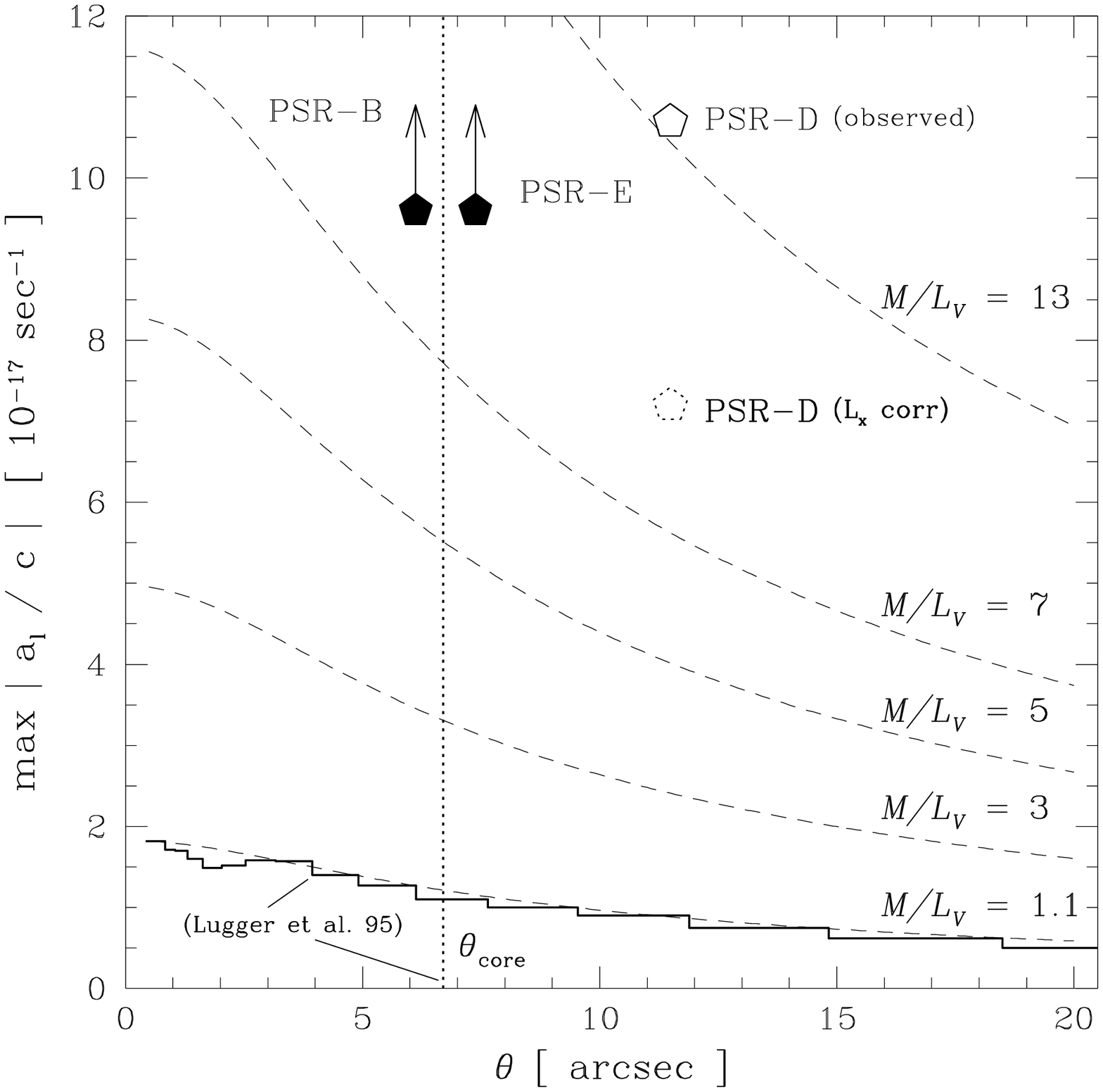} 
\figcaption[fig2.eps]{\label{fig:fig2}
\small{Maximum line-of-sight acceleration
$|a_{{l}_{max}}/c| = |\dot{P}/P|$ versus displacement $\theta$ with
respect to the center of NGC~6752.  The histogram represents the
prediction based on the available optical observations. The dashed
lines are analytical fits to the optical observations, labeled
according to the adopted mass-to-light ratio. The filled pentagons
represent lower limits to the line-of-sight acceleration based on the
observation of PSRs B and E. The open pentagon shows the value of
$\dot{P}/P$ for PSR-D assuming a negligible intrinsic $\dot{P}$.  If
various observed scalings (Becker \& Tr\"umper 1997;
Possenti et al. 2002; Grindlay et
al. 2002) between X-ray luminosity and spin-down power
for the MSPs are used to estimate the intrinsic $\dot{P}$, $\dot{P}/P$
for PSR-D assumes values in the interval between the open and the
dotted pentagon (see \S 3).}}
\vskip 0.5truecm
}

Assuming a typical mass-to-light ratio for these stars of 
$\sim 0.25$ (Phinney 1993), the number of
0.6--0.8~${\rm M_\odot}$ stars is $N_{\rm to}\sim 4\times
10^4$. 
Assuming a Salpeter initial mass function (IMF) $\propto
m^{-\alpha}$ with $\alpha=2.35$, the evolution of stars having initial
mass $\gapp 8~{\rm M_\odot}$ should have left at most $N_{\rm dr}\sim
4\times 10^3$ dark remnants. If they sank towards the cluster center
and are responsible for the high value of the mass-to-light ratio in
the core, their average mass would be ${\langle m_{\rm dr}\rangle}=
M_{cyl}(<\theta_{\perp,E})/N_{\rm dr} \gapp 3.2~{\rm M_\odot}$,
suggesting that at least some of them must be black holes.  The same
conclusions apply to any power-law IMF steeper than Salpeter's.
Conversely, if we assume that the low-luminosity objects within the
central $0.15$ pc are all 1.4\,${\rm M_\odot}$ neutron stars (NSs), this
would imply $\alpha_{\rm NS}<2.1$ for the slope of the IMF. Given the
very low upper limit on the density of any ionized gas, 
$\gapp 10^4$ NSs could reside in the inner region of NGC~6752 
without producing an accretion luminosity detectable in the deep 
{\it Chandra} exposure of Pooley et al. (2002).
Alternatively, if the population of the central region is
dominated by massive white dwarfs (descendents of stars of initial mass
$\gapp 2.5~{\rm M_\odot}$), we obtain a nominal limit
$\alpha_{\rm WD}<2.5$. However, these $\sim$1\,${\rm M_\odot}$ remnants
(less massive than the population of MSPs) should distribute out to larger
distances from the cluster center than PSR-E. A conservative 60\%
increase in their estimated number implies an IMF flatter than
Salpeter's in this case as well.

We note that both the uncertainties in the average mass-to-light ratio
of the 0.6--0.8~${\rm M_\odot}$ stars and the unknown number of these
stars which have been tidally stripped or evaporated from the cluster
since its formation could affect $N_{\rm to}$ and thus our
results. However, simulations of strongly concentrated clusters
(e.g.\ Joshi, Nave \& Rasio 2001\nocite{jnr01}) indicate that the
relative depletion of 0.6--0.8~${\rm M_\odot}$ stars should be smaller
than that of neutron stars, which escape the cluster potential well
due to supernova kicks at birth (Rappaport et
al. 2001)\nocite{r++01}. Such an effect would further increase the
average mass $\langle m_{\rm dr}\rangle$ of dark remnant stars and/or
make the IMF even flatter.

\section{Discussion}

Our results provide the first direct dynamical evidence for a high
density of unseen remnants in the core of a globular cluster,
suggesting a mass-to-light ratio $\gapp$10 in the core region. For
comparison, we note that a much smaller value $\sim 3,$ was obtained
for the case of M15 (Phinney 1993)\nocite{phi93}, a core-collapsed GC
long suspected to host a central black-hole (see e.g. van der Marel \&
Roeland 1999)\nocite{vr99}. The nature of the remnants in the core of
NGC~6752 is not clear as it strongly depends on assumptions about the
IMF. However, the possibility that they are black holes, or that many
stellar remnants have collapsed into a single massive black hole, is
intriguing.

In addition, the large offsets of PSRs A and C from the cluster center
indicate the occurrence of highly effective non-thermal dynamics in
the cluster core. No other GC shows an MSP ejected beyond its
half-mass radius (which requires a finely tuned impulse to avoid
prompt expulsion of the NS from the GC), while NGC~6752 has two of
them. They could result from exchange encounters in the core (Phinney
\& Sigurdsson 1991)\nocite{ps91} but the large offset of PSR-A
suggests also the occurrence of more powerful scattering events
(Colpi, Possenti \& Gualandris 2002)\nocite{cpg02}. The scattering
target must have treated this binary system as a point mass, ejecting
it without inducing appreciable eccentricity. Assuming a value of 1.4
M$_{\odot}$ for the pulsar, the total mass of the binary system
containing PSR-A is at least 1.6 M$_{\odot}$. Simple dynamical
considerations favor a scattering target significantly more massive
than the scattered binary, supporting the conclusion that it could be
of many solar masses. A black-hole binary system would be a natural
candidate. The {\it Chandra} X-ray observations do not place severe
limits on its maximum mass. The arguments of Grindlay et
al. (2001a\nocite{ghe+01a}) applied to an electron gas density $\lapp
0.025$ cm$^{-3}$ and a detection threshold of $\sim 10^{30}$ erg
s$^{-1}$, allow for up to a few hundred solar masses in the form of
black hole(s) present in the central region of NGC~6752.
 
Finally, the pattern of X-ray sources, roughly oriented midway between
the probable projected ejection directions of PSRs A and C, suggests a
preferential geometry for ejection, similar to that discussed by
Grindlay et al. (2001b) for NGC~6397. Its physical connection with
the inferred high density of unseen matter in the core is not clear,
but it could be a further signature of significant non-thermal
activity in the core region.

\acknowledgements {\small{We acknowledge stimulating discussions with
Luca Ciotti and Monica Colpi, and Piero Ranalli, Marcella Brusa and
Paolo Montegriffo for assistance with reduction of the X-ray data and
astrometry.  N.D'A. and A.P. received financial support from the
Italian Space Agency (ASI) and the Italian Minister of Research
(MIUR).  F.C. acknowledges support from NASA grants NAG5-9095 and
NAG5-9950. The Parkes radio telescope is part of the Australia
Telescope which is funded by the Commonwealth of Australia for
operation as a National Facility managed by CSIRO. The Chandra Data
Archive is operated for NASA by the SAO.}}

\begin{deluxetable}{llllll}
\tablewidth{0pt}
\tabcolsep 0.07truecm
\tablecaption{\label{tb:tpar} {Measured and Derived Parameters for Five Millisecond Pulsars in NGC 6752}}
\tablecolumns{6}
\tablehead{\colhead{Pulsar}&\colhead{PSR-A}&\colhead{PSR-B}&\colhead{PSR-C}&\colhead{PSR-D}&\colhead{PSR-E}}
\startdata
Name         &  J1911$-$5958A &  J1910$-$5959B & J1911$-$6000C  & J1910$-$5959D & J1910$-$5959E \\
R.A. (J2000) & 19$^{\rm h}$ 11$^{\rm m}$ 42\fs7562(2)
             & 19$^{\rm h}$ 10$^{\rm m}$ 52\fs050(4)  & 19$^{\rm h}$ 11$^{\rm m}$ 05\fs5561(7)
             & 19$^{\rm h}$ 10$^{\rm m}$ 52\fs417(2)  & 19$^{\rm h}$ 10$^{\rm m}$ 52\fs155(2) \\
Dec (J2000)  & $-$59$^\circ$ 58\arcmin 26\farcs900(2)
             & $-$59$^\circ$ 59\arcmin 00\farcs83(3)  & $-$60$^\circ$ 00\arcmin 59\farcs680(7)
             & $-$59$^\circ$ 59\arcmin 05\farcs45(2)  & $-$59$^\circ$ 59\arcmin 02\farcs09(2) \\
$P$ (ms)        &  3.2661865707911(5)
             &  8.35779850080(3)    &   5.277326932317(4)   
             &  9.03528524779(2)    &   4.571765939765(7)   \\
${\dot P}$   &    3.07(10)$\times 10^{-21}$
             & $-$7.99(5)$\times 10^{-19}$  &    2.2(7)$\times 10^{-21}$ 
             &    9.63(3)$\times 10^{-19}$  & $-$4.37(1)$\times 10^{-19}$   \\
Epoch (MJD)  &  51920.0     &  52000.0     &  51910.0     &  51910.0    & 51910.0     \\
DM (cm$^{-3}$pc) &   33.68(1)
                 &   33.28(4)       &         33.21(4)
                 &   33.32(5)       &         33.29(5)      \\
$P_{\rm orb}$ (d)   &  0.837113476(1) & \nodata & \nodata & \nodata & \nodata \\
${a_{\rm p} \sin(i)/c}$ (s)&  1.206045(2) & \nodata & \nodata & \nodata & \nodata \\
$T_{\rm asc}$ (MJD) & 51919.2064780(3)& \nodata & \nodata & \nodata & \nodata \\
Eccentricity        &  $<$ 10$^{-5}$  & \nodata & \nodata & \nodata & \nodata \\
M$_{\rm c}$ (M$_{\sun}$)  &  $>$ 0.19       & \nodata & \nodata & \nodata & \nodata \\
MJD Range   &  51710--52200    &  51745--52202    &  51710--52201 &  51744--52197  & 51744--52201 \\
No. TOAs    &  74          &  27          &  94          &  38         & 38          \\
residual ($\mu$s)     &   10            &  83     &  55     &  55     &  60     \\
$S_{1400}$ (mJy)      &   0.22      & 0.06    & 0.30    & 0.07    & 0.09    \\
Offset$^{\rm a}$($\arcmin$) & $6\farcm39$  & $0\farcm10$ &  $2\farcm70$ & $0\farcm19$ & $0\farcm13$ \\
\enddata
\tablenotetext{a}{Angular separation in the
plane of the sky between the MSP and the center of NGC~6752 (Harris 1996).}
\end{deluxetable}

\end{document}